\documentclass[preprint2]{aastex}

\usepackage{graphicx}
\usepackage{subfigure}
\usepackage{epstopdf}
\usepackage{url}

\shorttitle{The rings of Chariklo under close encounters with the giant planets}
\shortauthors{Araujo, R.A.N.; Sfair, R. \& Winter, O.C.}

\begin{document}

\title{The rings of Chariklo under close encounters with the giant planets}

\author{R.A.N. Araujo\altaffilmark{1}}
%\affil{Univ - Universidade Estadual Paulista, Grupo de Dinamica Orbital e Planetologia, CEP 12516-410, Guaratingueta, SP, Brazil.}
\email{ran.araujo@gmail.com}

\author{R. Sfair\altaffilmark{1}}
%\affil{Univ - Universidade Estadual Paulista, Grupo de Dinamica Orbital e Planetologia, CEP 12516-410, Guaratingueta, SP, Brazil.}
\email{rsfair@feg.unesp.br}

\and

\author{O.C. Winter\altaffilmark{1}}
%\affil{Univ - Universidade Estadual Paulista, Grupo de Dinamica Orbital e Planetologia, CEP 12516-410, Guaratingueta, SP, Brazil.}
\email{ocwinter@gmail.com}

\altaffiltext{1}{Univ - Universidade Estadual Paulista, Grupo de Din\^amica Orbital e Planetologia, CEP 12516-410, Guaratingueta, SP, Brazil.}

\begin{abstract}
The Centaur population is composed by minor bodies wandering between the giant planets and that frequently 
 perform close gravitational encounters with these planets, which leads to a chaotic orbital evolution.
 Recently, the discovery of two well-defined narrow rings was announced around the Centaur 10199 Chariklo.
 The rings are assumed to be in the equatorial plane of Chariklo and to
 have circular orbits. The existence a well-defined system of rings around a body in such perturbed orbital region
 poses an interesting new problem. 
 Are the rings of Chariklo stable when perturbed by close gravitational encounters with the giant 
 planets? 
 Our approach to address this question consisted of 
 forward and backward numerical simulations of 729 clones of Chariklo, with similar initial orbits, for a period of $100$ Myrs. 
 We found, on average, 
 that each clone suffers  along 
 its lifetime more than 150 close encounters with the giant planets within one Hill radius of the planet in question. 
 We identified some extreme close encounters able to significantly disrupt or to disturb the rings of Chariklo. 
 About $3~\%$ of the clones lose the rings and about $4~\%$ of the clones have the ring significantly disturbed.
 Therefore, our results show that in most of the cases (more than $90~\%$) the close encounters with the giant planets 
 do not affect the stability of the rings in Chariklo-like systems. Thus, if there is an efficient mechanism that creates the rings, then these structures may be common 
 among these kinds of Centaurs. 
\end{abstract}



%%%%%%%%%%%%%%%%%%%%%%%%%%%%%%%%%%%%%%%%%%%%%%%%%%%%%%%%%%%%%%%%%%%%%%%%%%%%%%%%%%%%%%%%%%%%%%%%%%%%%%%%%%%%%%%%%%%%%%%%%%%%%%%%%%%%%%%%%%%%%%%%%%%%%%%%%%%%%%%%%%%%%%%%%%%%%%%
\section{Introduction}
\label{sec_introduction}

\begin{deluxetable}{c c c c c c}
\label{T-chariklo}
%\rotate
\tabletypesize{\footnotesize}
\tablecolumns{6}
\tablewidth{0pc}
\tablecaption{Orbital and physical parameters of Chariklo and its rings}
\startdata
\hline
\hline
\multicolumn{2}{c}{Chariklo} &\colhead{}     &
\multicolumn{3}{c}{Rings} \\
\cline{1-2} \cline{4-6} \\
    a$^{(1)}$ 				& $15.74~au$  		& 	&		 			&$R1$	 				&$R2$			\\ \cline{4-6}
    e$^{(1)}$				& 0.171				& 	&Orbital radii (km)$^{(2)}$	&$390.6\pm3.3$				&$404.8\pm3.3$	\\ 
    i$^{(1)}$			& 23.4$^\circ$			& 	&Width (km)$^{(2)}$		&$7.17\pm0.14$				&$3.4^{+1.1}_{-1.4}$\\ \cline{4-6}
    Equivalent radius (km)$^{(2)}$ 	& 124 				& 					&Radial separation (km)$^{(2)}	$&\hspace{2.0cm}$14.2\pm0.2$	&				\\
    Mass (kg)$^{(3)}$	& $7.986\times10^{18}$		& 	&Gap between rings (km)$^{(2)}$	&\hspace{2.0cm}$8.7\pm0.4$		&				\\
\hline
    \multicolumn{6}{l}{$^{(1)}$Orbital elements obtained from JPL's Horizons system for the epoch MJD 56541. According to JPL the uncertainties} \\
    \multicolumn{6}{l}{in $a$, $e$ and $i$ are of the order $10^{-5}$, $10^{-6}$ and $10^{-5}$, respectively.} \\
    \multicolumn{6}{l}{$^{(2)}$ \cite{b1}.}\\ 
    \multicolumn{6}{l}{$^{(3)}$Calculated considering a density of $1~g/cm^3$, the equivalent radius of Chariklo and a spherical body.}\\
  \enddata
\end{deluxetable}

Among the orbits of the giant planets there is a population of small objects called Centaurs. 

There is not a consensus on the definition of the Centaur population. According to the Minor Planet Center - MPC/IAU, Centaurs are
celestial bodies with perihelion beyond the orbit of Jupiter and with semi-major axes smaller than the semi-major axis of Neptune \citep{b16}. 
Similarly, the JPL/NASA defines the Centaurs population as the objects with semi-major axes between $5.5$ au and $30.1$ au \citep{b17}.
\cite{b21} classify Centaurs as celestial bodies with orbits mostly in the region between Jupiter and Neptune and that typically 
cross the orbits of the giant planets. A similar definition is also found in \cite{b8}.

2060 Chiron was the first observed body of this population \citep{kowal79}.  
Since then, the  number of known Centaurs has grown. Currently, more than 400 objects are cataloged \citep{b10}, and from the flux of short period 
comets \cite{b6} estimated the total number of objects with diameter $>1$~km to be approximately $44000$.

The Centaur 10199 Chariklo was discovered in 1997 by the Spacewatch program \citep{b11}. 
In 2013, a stellar occultation revealed the existence of symmetric features encircling Chariklo, 
the second largest known Centaur. 
\cite {b1}, showed that these structures are a system of two narrow and well-defined 
rings. These discovery made the debut of minor bodies within the select group of ringed objects.

In that work, the authors estimated that the rings have orbital radii of approximately $391~km$ and $405~km$, and width of $7~km$ and $3~km$, respectively. 
They are assumed to be in the equatorial plane of Chariklo, with circular orbits. Besides an equivalent radius of 124~km derived 
from the same stellar occultation, few information 
about the physical properties of Chariklo is available. From the orbital position of the rings, 
they also estimated the density of the central body to be $1$~g/cm$^3$. 
Table~1 summarizes some of the physical and 
orbital parameters of Chariklo and its rings. 

\begin{figure*}
\centering
\mbox{
\subfigure[]{\includegraphics[scale=0.37]{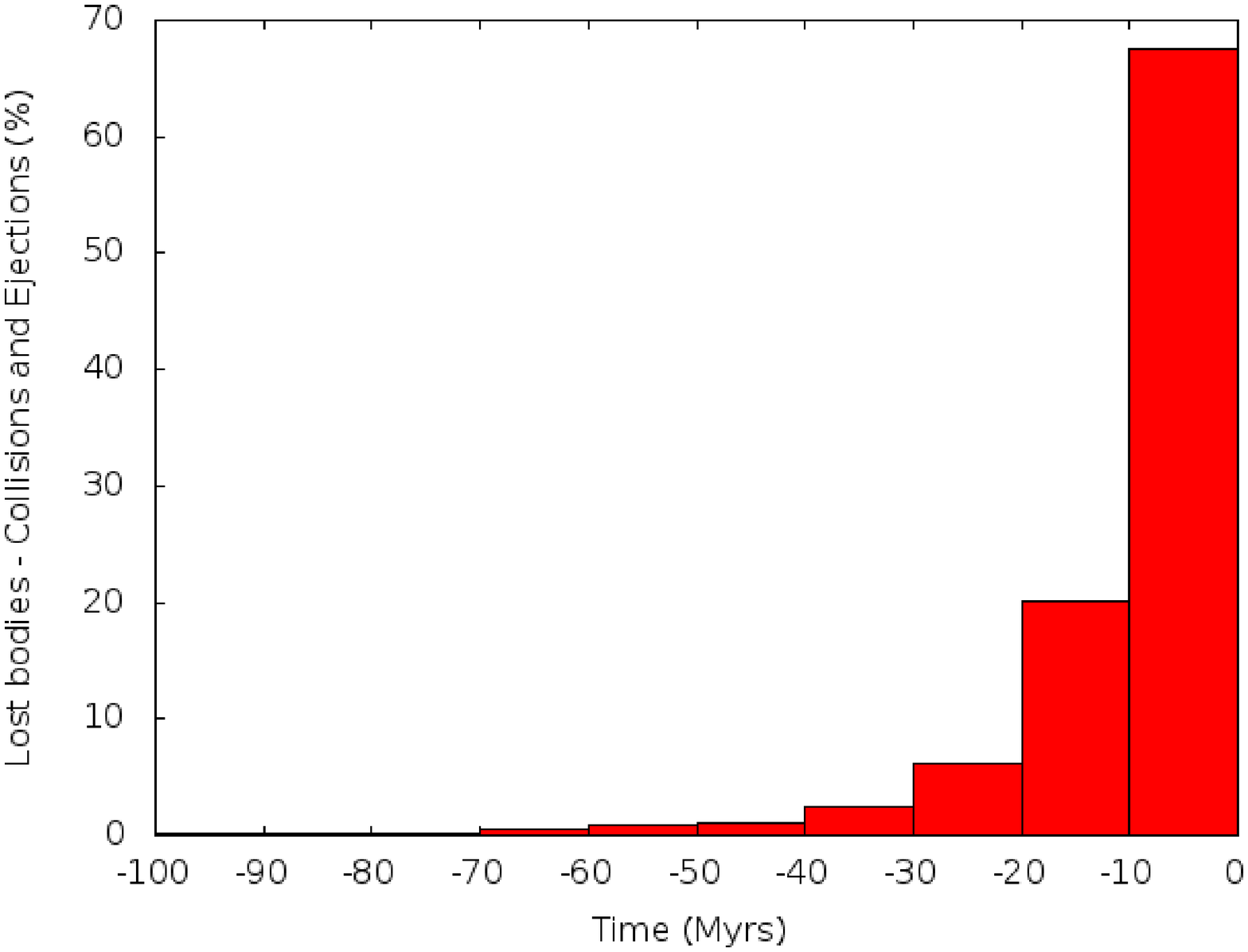}}\qquad
\hspace{-0.9cm}
\subfigure[]{\includegraphics[scale=0.49]{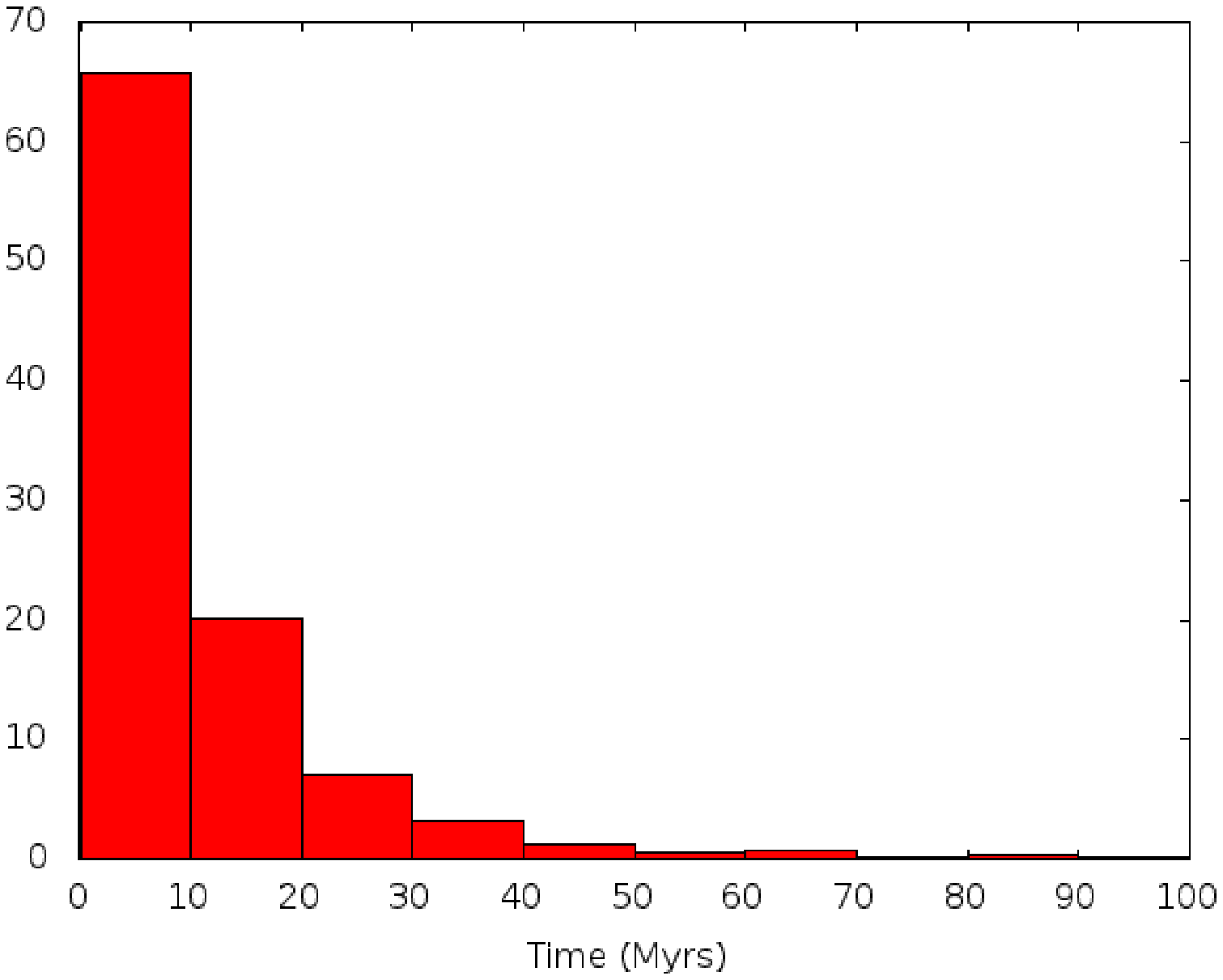}}} 
\caption{Histograms of the fraction of Chariklo clones lost within $100~Myrs`$, as a function of time.  a) Backward integration. b) Forward integration.
Throughout the numerical integration the clones could
be lost by ejection or collisions with one of the giant planets or with the Sun. Considering an ejection distance of $100~au$  
and with collisions defined by the physical radius of the planets and of the Sun.}
\label{fig_histogram}
\end{figure*}

A detailed study of the orbital evolution of the Centaurs was presented by \citet{b6}, 
where they analyzed the orbital evolution of 32 cataloged objects through numerical simulations 
of an ensemble of particles under the influence of the Jovian planets. They followed the particles both forward 
and backward in time and registered the dynamical evolution and fate of the particles. 
The orbital radius shows that Chariklo orbits between Saturn and Uranus, 
corresponding to a typical $U$ class object, i.e., those whose evolution is controlled by Uranus \citep{b15}. For Chariklo, 
they found the half-life to be 10.3~(9.68)~Myr.

The Centaur objects are transient. Therefore, a source is required to keep a steady state population. 
The idea of bodies coming from regions of the Solar System beyond Neptune and 
populating the region between the planets is well-accepted.

\cite{levison97} estimated, 
through numerical integrations, a number of $\approx 1.2\times10^7$ of comets transiting between the inner and outer Solar System originating from the Kuiper Belt.
In fact, \cite{b6}, estimated a flux of one body coming from the Kuiper belt and getting into the Centaurs population every 125 yr. 
\cite{b7}, present the Scattered Disk Objects - SDO (bodies with distance to the perihelion $q<30~au$ and semi-major axis $a>50~au$) 
as the most probable source of the Centaurs.
\cite{b18}, analyzed the role of the Oort cloud in determining the flux of cometary bodies
through the planetary system. They concluded that a substantial fraction of all known cometary bodies may have a 
source in the Oort cloud, including the Centaur population, which they defined as the population small bodies with perihelion $5<q<28~au$ and $a<1000~au$.
Following the same definition for Centaurs, \cite{b19} present that in fact more than $90\%$ of all Centaurs with $a > 60~au$, and $\approx50\%$ with $a<60~au$,
come from the Oort cloud.
In \cite{b20}, the Oort cloud is also pointed out as the source of Centaurs, especially those with high-inclination. In that work the Centaur population is defined 
as small bodies with perihelion between $15$ and $30~au$ and semi-major axis shorter than $100~au$. They showed that these objects probably were originated from the 
Oort cloud rather than the Kuiper belt or the scattered disc.

Along its orbital evolution a Centaur is strongly perturbed by the giant planets. \cite{b8}, illustrate in detail the effects of those perturbations 
on the orbit of five selected Centaurs. The close encounters with 
the giant planets are quite frequent. As consequence, the Centaurs present a characteristic chaotic orbital evolution.

The existence of a small body owning a well-defined system of circular rings within such perturbed 
population poses an interesting new problem. This scenario has motivated the development
of the present work. We investigated the stability of the rings of Chariklo when perturbed 
by close encounters with the giant planets. We analyzed how effective are the close encounters 
in disturb or disrupt the rings of Chariklo. 
Furthermore, the development of this study may allow us to quantitatively evaluate how propitious the region of the Centaurs is for such small bodies 
owning their own systems of rings. 
A brief qualitative discussion on this subject is presented by \cite{b12}, where it is proposed that the Centaur Chiron may also have rings. 
The possible existence of two small bodies belonging to the same population and owning a system of rings is quite interesting,
and indicate that such systems can be more frequent than expected.

Since our goal is to analyze the stability of the rings of Chariklo while perturbed by close encounter with the giant
planets, here we classify it as a Centaur while its orbit is mainly in the region between Jupiter and Neptune (as in \cite{b21}),
with maximum semi-major axis value of $a\leq50~au$. 

The structure of this paper is as follows. In Section \ref{sec_close_enc}, we present the initial conditions and the numerical method adopted in order to identify the 
close encounters of Chariklo with the giant planets. In Section \ref{sec_extreme}, we present the selection of the extreme close encounters, i.e., 
those encounters that could be capable of disrupting the rings of Chariklo. In Section \ref{simula_ring}, we describe how the rings were simulated along a close encounter. 
In Secs. \ref{sec_catastrophic},  \ref{sec_disturbing} and \ref{sec_ejec} we present the results, and in Section \ref{sec_final}, are our final comments and the major 
conclusions of the work.

\section{Close encounters with the giant planets}
\label{sec_close_enc}

The first step of the work consisted on selecting a representative sample of
close encounters performed by Chariklo with the giant planets. 

We considered a system composed by the Sun, the giant planets of the Solar System (Jupiter, Saturn, Uranus and Neptune), and a sample of 
clones, i.e., objects with the same mass and radius of Chariklo, but with small deviations on their orbits.

The clones were created following the procedure presented in \cite{b6}, where 729 clones were created from the original orbit assuming a 
variation of semi-major axis of $0.005~au$, a variation of eccentricity of $0.005$ and a variation of inclination of $0.01^{\circ}$.

The orbital elements of Chariklo and of the planets was obtained through JPL's Horizons system for the epoch MJD 56541. For the orbit 
of Chariklo at this epoch we have $a=~15.74~au$, $e=~0.171$ and $i=23.4^{\circ}$.

Considering these orbital elements, and taking the amplitude of variation as in \cite{b6} in such way that we have 729 clones, 
we created the clones of Chariklo orbiting the Sun as follows: $15.720\leqslant a\leqslant15.760~au$, taken every $0.005~au$; $0.151\leqslant e\leqslant0.191$, 
taken every $0.005$ and, $23.36^{\circ}\leqslant i\leqslant 23,44^{\circ} $, taken every $0.01^{\circ}$. 
The choice of these values resulted in nine values of semi-major axes, nine values of eccentricities and nine values of inclination.
The combination of these values resulted in 729 clones, each one with different values of $a$, $e$ and $i$.

Considering that Chariklo has an equivalent radius of $124$ km and a density of 1 $g/cm^3$ \citep{b1}, we estimated its mass as $M_{C}=~7.986\times10^{18}$ kg. 

We performed backward and forward numerical integrations of the system composed by the Sun, by the giant planets and by the clones, for a time span of 100 Myrs, 
using the adaptive time-step hybrid sympletic/Bulirsch-Stoer algorithm from \textsc{Mercury} \citep{b3}. 

Throughout the numerical integrations the clones did not interact with each other, but they could collide with the planets or escape from the system. 
The collisions were defined by the relative distance between the clones and the planets. If the clone-planet distance is smaller than the radius of the planet in question, 
then we have a collision. The physical radius of the planet is determined by \textsc{Mercury} assuming a spherical planet with uniform density.
We consider ejections as being the ejections from the Centaur population defined by the relative distance to the Sun of $100~au$. This value was adopted taking 
into account that if a clone reach the distance of $100~au$ and it is still in a elliptical orbit then necessarily the semi-major axis of the clone has to be 
greater than $50~au$, i.e, the clone is no longer classified as a Centaur, according to our definition.

As a result of the integrations, we see in the histograms in Figure \ref{fig_histogram} that more than $50\%$ of our sample was lost (ejections or collisions) in 10 Myrs, 
for both, backward and forward integrations.
These results show that the evolution of our sample is in agreement with the predicted evolution of the Centaurs, which has an estimated
mean lifetime of about 10 Myrs  \citep{b2}. We also note that there is a kind of symmetry on those results, which indicates that Chariklo is currently in the middle
of its median dynamical lifetime as a Centaur.

At the end of the forward integrations, we found that $\approx 94\%$ of the 729 clones were lost in the time span of 100 Myrs, being 683 clones lost by ejections and 
4 clones lost by collisions (three with Saturn and one with Jupiter). For the backward integration, we found that $\approx 99\%$ of the clones were
lost in 100 Myrs, being 719 ejections and 4 collisions (three with Jupiter and one with Saturn). 

Once we have characterized the evolution of the sample of clones as a whole, we then selected all close encounters of the clones within $1$ Hill radius with each giant 
planet performed within 10 Myrs (mean lifetime of the Centaurs). 
For this time span, there were registered 60159 close encounters for the forward integration, and
65293 encounters for the backward integration.
From Table~2 we see that in this case Uranus dominates, followed by Saturn, Jupiter and Neptune, for both, 
backward and forward integrations. This result is in agreement with works on the dynamics of Centaurs stating that the dynamics 
of bodies with similar orbits to the orbit of Chariklo should
be guided by Uranus, as discussed in \cite{b6}. 

However, we are interested in analyzing how the close encounters
of Chariklo with the giant planets might affect its rings. 
Therefore, we selected among all these registered close encounters those that are expected to perturb or disrupt
the rings. Following are the details of this analysis and the results obtained. 

\section{Extreme close encounters}
\label{sec_extreme}

\begin{deluxetable}{c c c c c c c c c c}
\label{tab_enc}
%\rotate
\tabletypesize{\footnotesize}
\tablecolumns{10}
\tablewidth{0pc}
\tablecaption{Registered close encounters of the clones with each one of the giant planets within $1$ Hill radius and within $1$ and $10$ rupture radius $(r_{td})$, 
in the time span of $10~Myrs$, for both, forward and backward integrations.}
\startdata
\hline
\hline
\multicolumn{1}{c}{}
 &&&  \multicolumn{3}{c}{Forward Encounters}
 && \multicolumn{3}{c}{Backward Encounters} \\ \cline{4-6} \cline{8-10}
Planet	 & Hill Radius$^{(1)}$   		&$r_{td}$ $^{(1)}$	& 			&  		  &  		     &	&	 	   	      	&	     	 &  \\
	 &(Planetary 	 			&(Planetary 		&1 Hill			&$r_{td}\leq 1$	  &$1<r_{td}\leq10$  &  &1 Hill      			&$r_{td}\leq 1$  &$1<r_{td}\leq10$ \\
	 &radii)				&radii)			&Radius$^{(2)}$	&		  &		     &  &Radius$^{(3)}$   	&		 &	\\	
\hline									
Jupiter	 &$\approx 740$	   			&$\approx 5$	     	&$16.6\%$  		&$5$		  &$36$		     &	&$18.3\%$       		&$5$    	 &$47$\\

Saturn	 &$\approx 1100$   			&$\approx 4$      	&$26.0\%$ 		&$1$		  &$34$		     &  &$24.2\%$	    		&$0$             &$25$\\

Uranus	 &$\approx 2700$   			&$\approx 5$      	&$48.0\%$  		&$0$		  &$18$		     &	&$46.9\%$	    		&$2$	         &$13$\\
	
Neptune	 &$\approx 4600$   			&$\approx 5$      	&$9.4\%$ 		&$0$		  &$2$		     &	&$10.6\%$	    		&$1$	         &$5$\\
\hline
\multicolumn{9}{l}{$^{(1)}$ The Hill radius and the rupture radius in terms of the radius of the planet in question.}\\
\multicolumn{7}{l}{$^{(2)}$Percentage relative to 60159 encounters.}\\
\multicolumn{7}{l}{$^{(3)}$Percentage relative to 65293 encounters.}
\enddata
\end{deluxetable}

In order to select the extreme close encounters, i.e., 
encounters with the giant planets that are expected to significantly affect the rings of Chariklo, we have calculated the Tidal Disruption Radius $(r_{td})$. 

According to \cite{b4}, the distance of the close encounter at which the tidal disruption of a binary may occur, is given by:
\begin{equation}
 r_{td} \approx a_{B}\left(\frac{3M_{Pl}}{M_1+M_2}\right)^{1/3}
 \label{eq_rtd}
\end{equation}
where $a_B$ is the semi-major axis of the binary, $M_{Pl}$ is the mass of the planet and $M_1$ and $M_2$ are the masses of the components of the binary. 

For a particle orbiting Chariklo with $a_B=410$ km (approximated outer limit of the ring), we calculated the $r_{td}$ for encounters
performed with each one of the giant planets. These values are presented in Table~2.

It is important to point out that this is an approximated value since it does not take into account the relative velocity of the bodies
at the moment of the encounter. \cite{b13}, has shown that not just the distance of the encounter, but also the relative encounter velocity 
determine how significantly a body will be disturbed by a close encounter. Such effect were also discussed by \cite{b14}, when they compared their numerical
analysis on the disruption of NEAs binaries due to close encounters with the Earth, with the analytical prediction given by Eq. \ref{eq_rtd}, 
showing the dependence of the results on the relative velocity of the encounters.

Nevertheless, for our purposes the approach given by Eq. \ref{eq_rtd} is adequate. 
Knowing that this value is an approximation, we then selected among all the registered close encounters those that had the minimum distance within $10~r_{td}$. 
For the forward integration, we see  that most of them (about 3/4) occurred with Jupiter and Saturn (Table~2).
Very few encounters occurred within 1$r_{td}$ (none with Uranus or Neptune).
For the backward integration, we see that the extreme encounters with Jupiter and Saturn still prevail, but
here we note the occurrence of a few encounters occurred within 1$r_{td}$ with Uranus and Neptune.

We explored the effects of each one of these extreme encounters ($\leq 10r_{td}$) on the particles of Chariklo's rings, as follows.

\section{Simulating the rings}
\label{simula_ring}

\begin{deluxetable}{c c c c c c c c}
\label{table_rings}
%\rotate
\tabletypesize{\footnotesize}
\tablecolumns{8}
\tablewidth{0pc}
\tablecaption{Registered catastrophic and disturbing encounters of the rings of Chariklo due to extreme close encounters with each one of the giant planets, in the time span of 10 Myrs.}
\startdata
\hline
\hline
\multicolumn{1}{c}{}
&   \multicolumn{3}{c}{Forward Encounters}
 && \multicolumn{3}{c}{Backward Encounters} \\ \cline{2-4} \cline{6-8}
		&Catastrophic$^{(1)}$		&Disturbing$^{(2)}$&Survival Time$^{(3)}$	&	 &Catastrophic$^{(1)}$		&Disturbing$^{(2)}$&Survival Time$^{(3)}$  		\\
		&				&			&[Max:Min](years)			&	 &				&		&[Max:Min](years)	\\
\cline{2-4} \cline{6-8}
Jupiter		&6				&9			&[16,125:221,739]    		&	 &6				&16			&[-49,245:-546,550]            \\
%		&				&			&			      &  &				&			& \\

Saturn		&4				&7			&[56,650:623,559]    		& 	 &0				&7 			& -                \\
%		&				&			&  			      &	 &				&			&		 \\
		
Uranus		&0				&0			&  -        		      	&	 &3				&1			&[-1,371,579:-4,401,849] \\
%		& 				& 			& 			      & 	 &  				& 			& \\	
Neptune		&0				&0			&  - 			      	&	 &1				&2			&-1,499,269      \\
%		&				&			&			      &	 &				&			& \\
\hline
\multicolumn{8}{l}{$^{(1)}$ Rings are completely removed.}\\
\multicolumn{8}{l}{$^{(2)}$ External particles removed but the rings survive.}\\
\multicolumn{8}{l}{$^{(3)}$ Minimum and maximum survival time registered among the clones after they suffered a catastrophic encounter.}\\
\multicolumn{8}{l}{excluding the immediate ejection cases.}\\

\enddata
\end{deluxetable}

At the second step of the work, we numerically simulated the extreme close encounters including massless particles around Chariklo.
 
\begin{figure}
\centering
\subfigure[]{\includegraphics[scale=0.35]{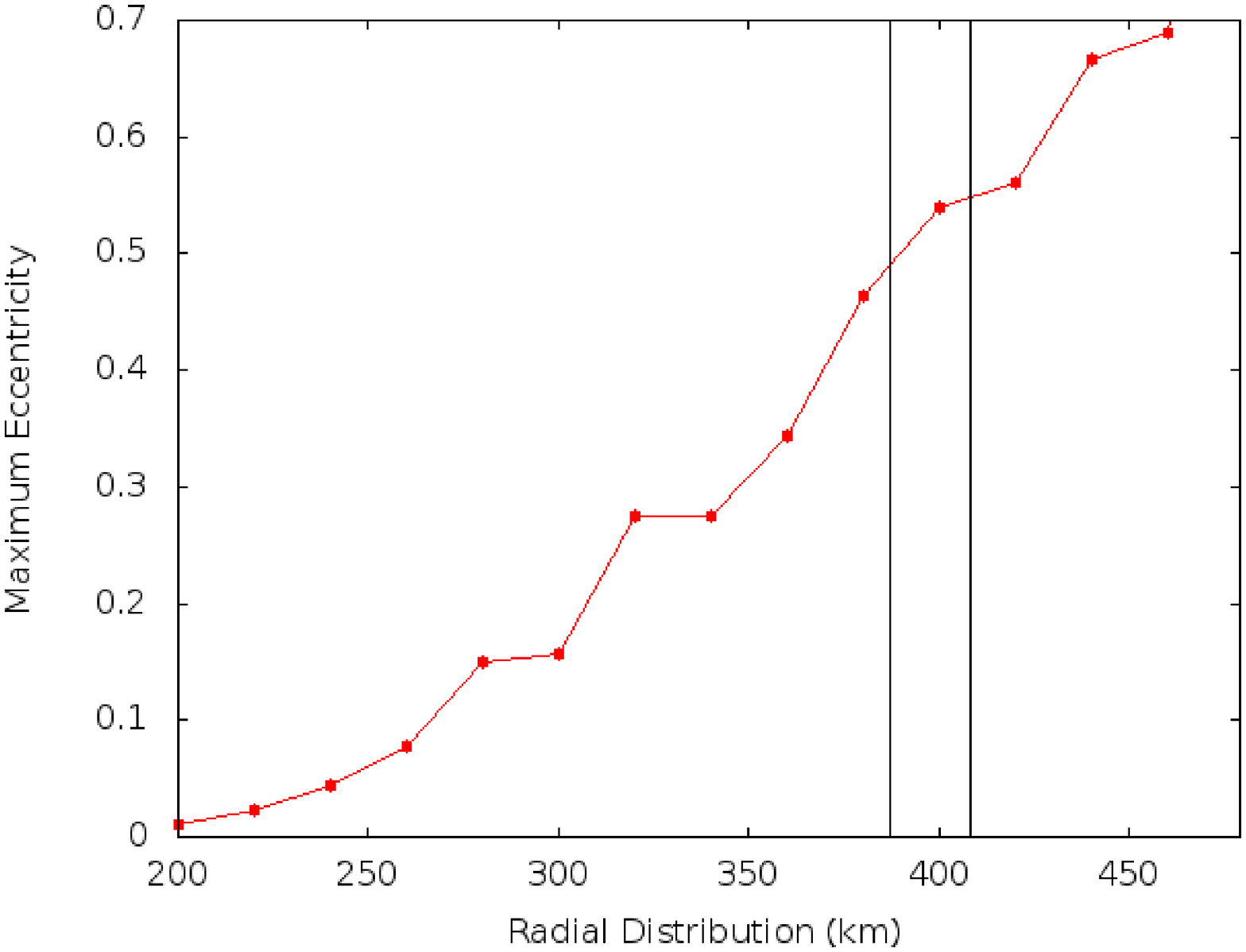}}
\subfigure[]{\includegraphics[scale=0.35]{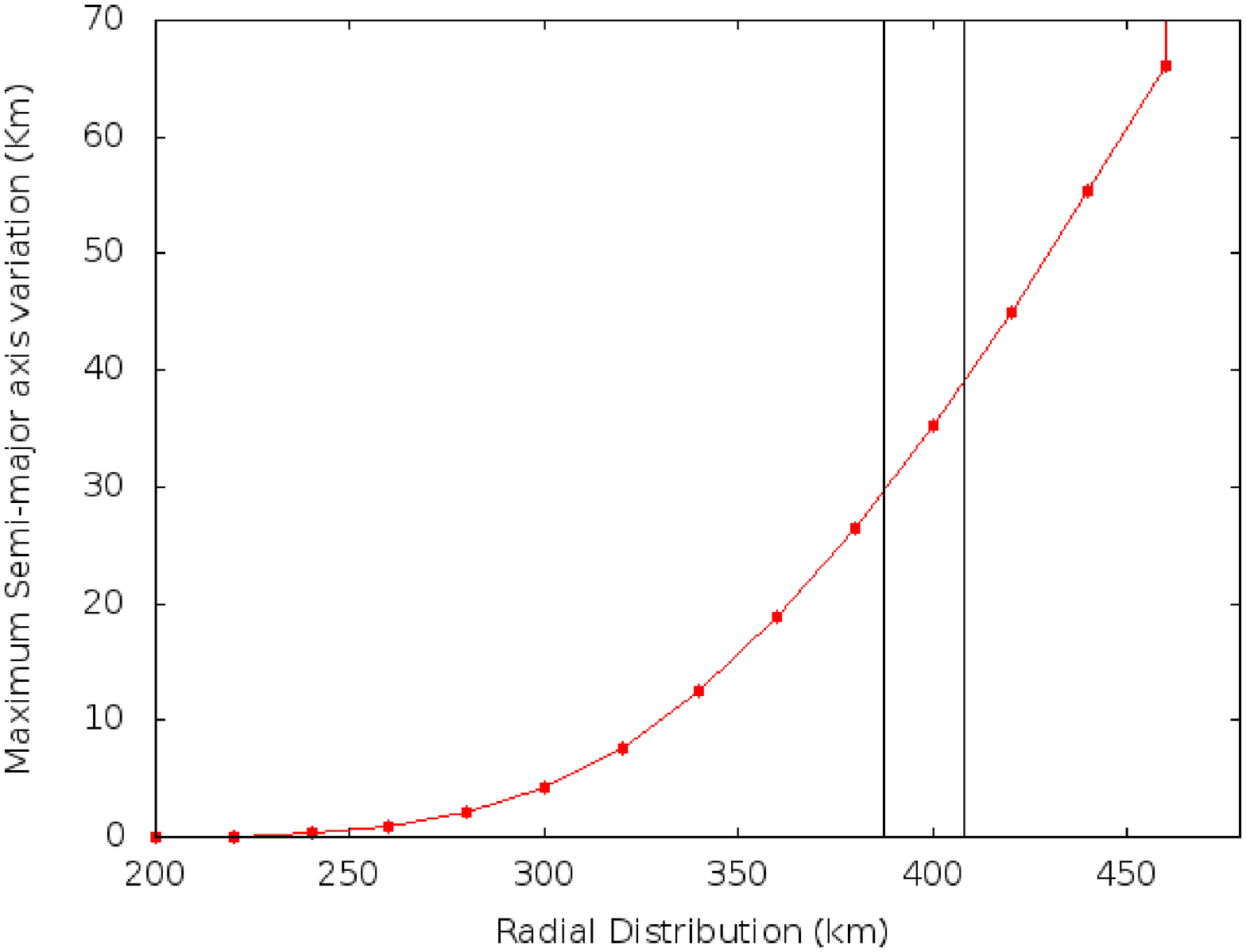}}
\caption{Example of a disturbing close encounter caused on the ring by a close encounter with Jupiter. 
a) Maximum final change in eccentricity. b) Maximum final change in semi-major axis (km).
The plots show the maximum final of those elements among the hundred particles that share the same initial radial
position. The vertical lines indicate the boundary of the rings' region. 
The encounter were performed with a minimum distance of 6.6 planet radius, with a relative velocity $V_{\infty} = 8.00$ km/s. The encounter
resulted in the critical radius of $R_C = 460$ km, meaning that no particle in the region of the rings were removed.}
\label{fig_2}
\end{figure}

According to Section \ref{sec_close_enc}, a close encounter is registered when a clone of Chariklo crosses the limit of $1$
Hill radius of any of the giant planets. At this crossing moment, we recorded the position and 
the velocity of these bodies, relative to the Sun. These values are the initial conditions for the simulations of the extreme close encounters, i.e., 
encounters with minimum distance within $10 r_{td}$.
Thus, at this step, the numerical simulations always involve the Sun, the bodies performing the close encounter (planet and Chariklo), and a sample of particles orbiting Chariklo.

We considered particles with circular equatorial orbits, radially distributed from 200 km to 1000 km,
taken every 20 km. For each radial distance there were considered 100 particles in a random angular distribution. 
Such combination of values resulted in a total of 4100 particles orbiting Chariklo.

The pole orientation of Chariklo was considered perpendicular to the orbital plane.
This is a reasonable approach since the rings are assumed to be in the equatorial plane of Chariklo, and we are 
interested in the maximized radial perturbations of the rings.

The encounters were simulated for a time span of 1 year using the adaptive time-step Gauss-Radau numerical integrator, keeping the accuracy  $10^{-12}$ \citep{b5}. 
Throughout the integration the particles could collide with Chariklo or be ejected. The collisions were defined by the equivalent radius of Chariklo $(124~km)$ .
The ejections were defined by the energy of the Two-body Problem Chariklo-particle.

According to Table~2, there were simulated 96 close encounters for the forward case, and 98 for the backward case.
The results of these simulations and their implications are presented in Table~3, and they are discussed in the following sections.

\section{Catastrophic Encounters}
\label{sec_catastrophic}

We classify as catastrophic those close encounters that lead to the complete removal of the particles in the region of the rings of Chariklo. 
Knowing that the particles of the rings are distributed in the range of $\approx390~km$ to $\approx405~km$, we defined that there was a catastrophic encounter
if at the end of our simulation particles distributed beyond 380 km were lost by ejection or collision as defined in Sec. \ref{simula_ring}.

The results presented in Table~3 show that in about $10\%$ of the simulations the rings were removed from Chariklo due to close gravitational encounters with the 
giant planets, for both, backward and forward integrations. 

For the forward integration, we found that only extreme encounters with Jupiter and Saturn were able to fully remove the rings.
For the backward integration there were a few cases where Uranus and Neptune were able to do that. 
Our data suggest that Uranus and Neptune might have influenced the existence of the rings in the past, 
but from now on Jupiter and Saturn would play this role. 
However, more simulations are required to investigate whether this is a real difference in the forward/backward 
evolution or the result of statistical artifact due to the small number of extreme close encounters.

In the cases of forward integration in time, after the removal of the rings, Chariklo remained as a Centaur for much less 
than a million years, i.e., the rings were destroyed in the very last stage of Chariklo's orbital evolution among the giant planets.

In the cases of backward integration in time, the removal of the rings occurred much before Chariklo complete its first million years as a Centaur, 
i.e., in its first stage of orbital evolution among the giant planets.

\section{Disturbing Encounters}
\label{sec_disturbing}

\begin{deluxetable}{c c c c c}
\label{tab_coefic}
\tabletypesize{\footnotesize}
\tablecolumns{5}
\tablewidth{0pc}
\tablecaption{List of significant disturbing encounters $(\Delta e\geqslant0.01)$ with the giant planets for both, forward and backward integration within the time span of 10 Myrs. 
The immediate ejection cases were excluded.}
\startdata
\hline
\hline
\multicolumn{5}{c}{Jupiter - Forward}\\
\hline
Minimum encounter 		&$V_{\infty}$ 	&Maximum  		        & Maximum  		&Survival	\\
distance			&(km/s) 	& ring	        		& ring 			&Time$^{**}$	\\
(Planet radius)$^*$		&		&$\Delta a$ (km)		&$\Delta e$ 		&  (years)		\\
\hline
%6.0			&7.34		      	&3.08			&0.060			&		\\	
6.6			&4.00			&35.28			&0.54			&451,203		\\
6.9			&4.67			&7.61			&0.28			&509,333		\\
8.5			&5.81			&34.40			&0.45			&43,868		\\
8.8			&3.76			&28.20			&0.32			&6,891		\\
9.2			&5.03			&44.33			&0.57			&1,140,000		\\
15.4			&7.00			&0.04			&0.02			&55,034		\\
16.0			&7.78			&0.02			&0.01			&59,483		\\
\hline
\multicolumn{5}{c}{Saturn - Forward}\\
\hline
%Minimum 		&$V_{\infty}$	&Maximum		&Maximum	&Survival		\\
%approximation		&(km/s)  	&ring  			&ring		&Time* 	 	\\
%(Planet radius)		&		&$\Delta a$ (km)	&$\Delta e$	&(years)			\\
6.8			&4.87		      	&24.77		&0.34		&22,115			\\	
7.4			&3.12			&30.13		&0.53		&337,731			\\
8.2			&4.42			&1.23		&0.07		&87,594			\\
9.3			&5.29			&0.68		&0.06		&253,520			\\
10.4			&3.63			&0.14		&0.031		&573,463			\\
\hline
\multicolumn{5}{c}{Jupiter - Backward}\\
\hline
%Minimum 		&$V_{\infty}$ 		&Maximum  		& Maximum  		&Survival	\\
% approximation		&(km/s) 		& ring	        	& ring 			&Time*	\\
%(Planet radius)		&			&$\Delta a$ (km)	&$\Delta e$ 		&  (years)		\\
4.2			&9.87			&25.59			&0.20			&-217,796		\\	
8.3			&4.17			&17.34			&0.40			&-18,259		\\
8.4			&4.96			&13.63			&0.33			&-23,770		\\
8.4			&10.32			&65.85			&0.51			&-8,204		\\
8.5			&7.23			&2.42			&0.20			&-280,371		\\
11.1			&4.93			&0.065			&0.03			&-84,066		\\
11.1			&5.20			&5.74			&0.19			&-153,800	\\
11.9			&6.62			&2.33			&0.10			&-537,024		\\
13.4			&3.17			&0.28			&0.04			&-309,943		\\
13.5			&7.40			&0.06			&0.02			&-36,896		\\
14.4			&1.25			&0.03			&0.01			&-98,210		\\
15.4			&5.72			&0.02			&0.01			&-45,522		\\
16.7			&9.65			&0.03			&0.01			&-1,634,011		\\			
\hline
\multicolumn{5}{c}{Saturn - Backward}\\
\hline
%Minimum 		&$V_{\infty}$		&Maximum		&Maximum	&Survival		\\
%approximation		&(km/s)  		&ring  			&ring		&Time* 	 	\\
%(Planet radius)		&			&$\Delta a$ (km)	&$\Delta e$	&(years)			\\
%5.9			&2.73		      	&0.0013			&0.0000		&-143,122			\\	
%7.2			&3.60			&0.0014			&0.0000		&-103,462			\\
7.9			&2.68			&13.79			&0.300		&-405,570			\\
%8.2			&5.90			&0.01			&0.002		&-63,707			\\
%8.4			&3.68			&0.0021			&0.0000		&-110,002			\\
%10.1			&4.75			&0.0021			&0.0000		&-191,165			\\
\hline
\multicolumn{5}{c}{Uranus - Backward}\\
\hline
%Minimum 		&$V_{\infty}$		&Maximum		&Maximum	&Survival		\\
%approximation		&(km/s)  		&ring  			&ring		&Time* 	 	\\
%(Planet radius)		&			&$\Delta a$ (km)	&$\Delta e$	&(years)			\\
10.8			&3.00		      	&3.73			&0.11		&-2,504,843			\\	
\hline
%\multicolumn{5}{c}{Neptune - Backward}\\
%\hline
%Minimum 		&$V_{\infty}$		&Maximum		&Maximum	&Survival		\\
%approximation		&(km/s)  		&ring  			&ring		&Time* 	 	\\
%(Planet radius)		&			&$\Delta a$ (km)	&$\Delta e$	&(years)			\\
%7.3			&2.44		      	&0.0028			&0.00		&-4,014,495			\\	
%\hline
\multicolumn{5}{l}{$^{*}$ The minimum encounter distance given in terms of the radius of the planet in question.}\\
\multicolumn{5}{l}{$^{**}$ Survival time after a disturbing encounter.}\\
\enddata
\end{deluxetable}

There were cases in which the particles of the rings were not removed, but their orbits were significantly 
changed due to a significant perturbation caused by the encounter with a giant planet.

In Figure 2, we present an example of the effects of such kind of encounter with Jupiter. 
The plots show the maximum final variation in semi-major axis and eccentricity among the hundred particles that share the same
initial radial position.
As expected, the values increase with the radial distance. 
In the region of the rings (indicated by the vertical lines) the semi-major axis changes by more than 30 km and the eccentricity 
grows up to more than 0.5.

The results of the encounters that produced at least some noticeable variations $(\Delta~e\geqslant0.01)$ on the orbits of the rings
are presented in Table~4, for backward and forward integrations, excluding the encounters followed by ejection.

For the forward integration, we observed only five encounters with Jupiter and two with Saturn that resulted in eccentricities larger than 0.1 
for the ring particles.
None of the encounters with Uranus and Neptune showed significant orbital variation on the particles of the rings. Similarly, if we look backward, 
we see fewer close encounters that could have increased the eccentricity of the particles of the rings for values larger than 0.1 (5 with Jupiter, 1 with Saturn and 
1 with Uranus).

Since Chariklo, as observed now, has well-defined narrow circular rings \citep{b1}, it might mean that it did not suffer any of those disturbing encounters 
or if it happened it was so long ago so that the rings had time to evolve damping the eccentricity and reshaping them. Both possibilities are compatible
with our results, and as shown in Table~3, the probability of catastrophic or disturbing encounters is very low, and probably occurred several million 
years ago.

\section{The ejection cases}
\label{sec_ejec}
The close encounters of Chariklo with any of the giant planets followed by ejections are especially interesting.
If we look forward, those would be the encounters responsible for removing Chariklo from the population of Centaurs. On the other
hand, looking backward, those events would be the ones that brought Chariklo into the region that we defined as the Centaurs population, according to our definition 
(see Sec. \ref{sec_introduction}). We then looked for the last close encounter (within $1$ Hill radius) that every clone suffered before being ejected.

For the forward integration, we found that approximately $36\%$ of the clones suffered
an encounter with Jupiter before the ejection, $45\%$ with Saturn, $10\%$ with Neptune, 
and $9\%$ with Uranus. 

For the backward
integration, we found that approximately $27\%$ of the encounters were with Jupiter,
$42\%$ with Saturn, $20\%$ with Neptune, and $11\%$ with Uranus. 

As expected, the same pattern is obtained when we restrict our analysis to the total of extreme close encounters, i.e., encounters
performed within $1 \leq r_{td} \leq 10$ (Table~2 and Table~3).
We found for the forward integrations that 
4 clones of Chariklo were ejected after an extreme close encounter with Jupiter, 3 with Saturn, and none with Uranus 
or Neptune. For the backward integrations, we registered 11 clones ejected after a close encounter with Jupiter, 5 with Saturn, 2 with Uranus, and
1 with Neptune.  

Among these encounters, we found that only in 6 cases Chariklo was ejected just by the close encounter that disrupted the rings. 
For the forward integrations, such kind of event occurred 2 times after an encounter with Jupiter and one time after an encounter with Saturn.
For the backward integrations, we registered 2 cases with Jupiter and one with Uranus. 

Thus, our results indicate that for both backward and forward integrations, Jupiter and Saturn are the mainly responsible for 
inserting and removing Chariklo from the Centaurs region.

However, it is important to point out that since our ejection criteria requires that the object goes beyond $100~au$, it is more likely to have 
ejections caused by stronger close encounters as those produced by Jupiter and Saturn. Another consequence of our ejection criteria is that there are 
bodies that became TNOs without being considered ejected. For example, a Plutino-like object with $a=39~au$ and $e=0.2$ would remain as a Centaur in our 
simulations.

Anyhow, regardless the way that Chariklo followed to get in or to go out from the Centaur region, our results allow us to conclude 
that it may have brought the rings when it was introduced into this region and that it can keep them while leaving it.

\section{Final Comments}
\label{sec_final}

In the present work we explored the dynamics of the Centaur 10199 Chariklo and the stability of its rings when disturbed by 
frequent close encounters with the giant planets.

Through numerical integrations we analyzed the orbital evolution of Chariklo while a Centaur (orbits mainly in the region between 
Jupiter and Neptune and with $a \leq 50$ au). 

To do so we considered a sample of 729 clones of Chariklo with small deviations
on semi-major-axis, eccentricity and inclination. 
 
Throughout the lifetime of Chariklo in the population of the Centaurs, we found that the close encounters within $1$ Hill radius 
with Uranus are more frequent than the encounters with the other planets. Nevertheless, if we look to the most significative close encounters, i.e, those 
able to disturb or disrupt the rings of Chariklo, we found that Jupiter and Saturn dominate. 

For the forward integrations the most significative encounters happened in the last stage of Chariklo
among the giant planets, and they were performed exclusively with Jupiter and Saturn. This indicates that
Uranus and Neptune may not play an important role in the future dynamics of the rings of Chariklo, but these planets 
may have had more influence on the rings in the past. 
For the backward integrations we found few cases of catastrophic encounters of Chariklo with these planets, being three catastrophic
encounters with Uranus and one with Neptune. Nevertheless, the difference between forward and backward integrations is subtle and
probably arised from a statistical analysis based on a small number of extreme close encounters.

In total, for both backward and forward integrations, we found that the number of close encounters really able to completely 
disrupt the rings of Chariklo is small
($\approx~3 \%$ of the clones),  and they most probably happen in the beginning or in 
the final stages of Chariklo's time as a Centaur.

Thus, although a typical Centaur such as Chariklo present a chaotic orbital evolution \citep{b6}, we found that 
if the body had those rings before becoming a Centaur or acquired them in the region among the giant planets, 
then these rings will probably survive throughout its Centaur life. Hence, the formation of the rings of Chariklo while a 
Centaur is not mandatory.

Therefore, our major conclusion is that Centaurs experience a propitious environment to the
existence of rings. Furthermore, if there is an efficient mechanism that creates the rings, then these structures may be common among 
the bodies of this population.

One possible mechanism to create rings around Centaurs could be the outcome of extreme close encounters with the giant planets obtained here. 
This possibility will be analyzed in a future work.

\section{Acknowledgments}
This work was funded by CAPES, CNPq and FAPESP (proc. 2011/08171-3).  This support is
gratefully acknowledged. The authors are also grateful to the
anonymous referee for suggestions that significantly improved the
paper.

%++++++++++++++++  REFERENCES ++++++++++++++++++++
\renewcommand{\refname}{REFERENCES}

%\label{lastpage}
\end{document}